\documentclass[pre,aps,twocolumn,superscriptaddress]{revtex4-1}
\pdfoutput=1
\usepackage{amssymb}
\usepackage{epsfig}
\usepackage{amsmath}
\usepackage{subfigure}
\usepackage{graphicx}
\usepackage{textcomp}
\usepackage{url}
\usepackage{float}
\usepackage{color}
\usepackage{cases}
\usepackage[normalem]{ulem}

\usepackage[titletoc,title]{appendix}

\usepackage[colorlinks=true, urlcolor=blue, anchorcolor=blue, citecolor=blue,filecolor=blue,linkcolor=blue,menucolor=blue
]{hyperref}

\usepackage{color}

\begin{document}
\title{Anomalous scaling of dynamical large deviations of stationary Gaussian processes}

\author{Baruch Meerson}
\affiliation{Racah Institute of Physics, Hebrew University of
Jerusalem, Jerusalem 91904, Israel}

\begin{abstract}
Employing the optimal fluctuation method (OFM), we study the large deviation function of long-time averages $(1/T)\int_{-T/2}^{T/2} x^n(t) dt$, $n=1,2, \dots$, of centered stationary Gaussian processes. These processes are correlated and, in general, non-Markovian. We show that the anomalous scaling with time of the large-deviation function, recently observed for $n>2$ for the particular case of the Ornstein-Uhlenbeck process, holds for a whole class of stationary Gaussian processes.
\end{abstract}

\maketitle
\nopagebreak

\section{Introduction}
\label{intro}

Large deviations of stochastic processes remain a focus of non-equilibrium statistical mechanics
and probability theory \cite{Varadhan,O1989,DZ,Hollander,T2009,MS2017,Touchette2018}. Among them an
important role is played by \emph{dynamical}, or additive large deviations: of quantities, obtained by integrating the stochastic process or a function of it (and/or of its time derivative) over time.  Dynamical
large deviations emphasize temporal correlations of the process and exhibit a non-equilibrium behavior even if the system is in equilibrium. Here we consider long-time averages of positive integer powers of some centered stationary stochastic processes in continuous time. If we denote such a process by  $x(t)$, then the time-average $\bar{x^n}$ is defined by
\begin{equation}\label{xkbar}
\bar{x^n} = \frac{1}{T} \int_{-T/2}^{T/2} x^n(t)\,dt, \quad n=1,2, \dots .
\end{equation}
Obviously, $\bar{x^n}$ is a random quantity. What is the probability $\mathcal{P}(a,T)$ of observing a specified value $\bar{x^n} = a$, when the averaging time $T$ is much longer than the characteristic correlation time  $\tau$ of the process? For many stationary \emph{Markov} processes, the logarithm of $\mathcal{P}(a,T)$  turns out to be proportional to $T$ at large $T$:
\begin{equation}\label{normal}
-\ln \mathcal{P}(a,T\to \infty) \simeq T f(a) .
\end{equation}
This simple scaling behavior of $\ln \mathcal{P}(a,T)$ is considered ``normal". In this case the rate function $f(a)$ can be obtained from the
largest eigenvalue of the Feynman-Kac equation for
the generating function of $\bar{x^n}$.
For a whole class of models it is possible, and convenient, to recast the eigenvalue problem into a problem of finding the ground state energy of
an effective quantum oscillator \cite{MajumdarBray}, see also Ref. \cite{Touchette2018} for an accessible review.

Remarkably, Nickelsen and Touchette (NT) \cite{Nickelsen} have recently observed that the scaling behavior of $\ln \mathcal{P}(a,T)$ can be anomalous. For large $T$ and large $a$ they obtained
\begin{equation}\label{anomalous}
-\ln \mathcal{P}(a\to \infty,T\to \infty) \simeq T^{\xi} f(a),
\end{equation}
with $\xi<1$. NT considered the Ornstein-Uhlenbeck (OU) process -- a Markov process, generated by the Langevin equation
\begin{equation}\label{OULangevin}
 \dot{x}=- \frac{x}{\tau}+\epsilon \,\eta(t),
\end{equation}
where $\eta(t)$ is a Gaussian white noise, and $\langle\eta(t) \eta(t')\rangle = \delta(t-t')$. At long times,
this process is stationary with the covariance
\begin{equation}\label{OUcorrelator}
\langle x(t) x(t^\prime)\rangle = \frac{\epsilon^2\tau}{2}\,e^{-\frac{|t-t^\prime|}{\tau}}.
\end{equation}
For $n=1$ and $2$ the Feynman-Kac equation for this problem leads to a Schr\"{o}dinger equation with a quadratic potential.  The normal scaling immediately follows \cite{MajumdarBray,Touchette2018}. For $n>2$ the effective quantum potential is not confining, implying a breakdown of the standard dominant-eigenvalue formalism, and raising the possibility of a different scaling behavior of $\ln \mathcal{P}(a,T)$ with $T$.  In order to probe this regime, NT employed the optimal fluctuation method (OFM) (sometimes also called the weak noise theory) \cite{Onsager,Freidlin,Dykman,Graham}.  In the OFM the problem reduces to a minimization of the action functional of the OU process, where the constraint  $\bar{x^n} = a$ is accommodated via a Lagrange multiplier. For the OU process the minimization procedure defines an effective one-dimensional classical mechanics, and the dominant contribution to  the mechanical action [that is, to  $-\ln \mathcal{P}(a,T)]$ comes from the \emph{optimal path}  $x(t)$ - the solution of the minimization problem constrained by the condition $\bar{x^n} = a$.

As NT found, for $n=1$ and $2$ and in the regime $T\gg \tau$, the optimal path $x(t)$ stays, for most of the time,  very close to the unique stable fixed point on the phase plane of the effective classical mechanics.  (An identical behavior was previously predicted, by a different version of the OFM \cite{MZ2018}, for the continuous-time Ehrenfest urn model \cite{EUM} and its extensions.)  As a result, the classical action $S\simeq T f(a)$ is proportional to $T$,  immediately leading to Eq.~(\ref{normal}).

For $n>2$ the stable fixed point on the phase plane continues to exist. However, an additional solution for the optimal path $x(t)$ appears \cite{Nickelsen}.  This solution is localized in time on a time scale of $\tau$. As $T\to \infty$ the localized optimal path becomes a homoclinic orbit encircling the stable fixed point on the phase plane. The localized solution has a lesser action, and it causes the anomalous scaling (\ref{anomalous}) with $\xi=2/n$ \cite{Nickelsen}.

Scaling behaviors, different from the ``normal" scaling of the type (\ref{normal}), were  previously observed in the long-time statistics of time-integrated quantities in spatially extended systems such as diffusive lattice gases. These include the statistics of time-integrated current on an infinite line \cite{DG2009} and the statistics of the position of a tagged particle in the single-file diffusion \cite{tagged}.  The emergence of anomalous scaling in a  (much simpler) \emph{stochastic process}, that is in zero spatial dimension, caught us by surprise.

What is the ``warning signal" that points out to anomalous scaling? For a whole class of Markov processes this is a non-confining quantum potential. But what if the correlations make the process non-Markov, and the Feynman-Kac method does not apply? This is the question that we address in this work.   The OU process, that NT dealt with, is unique because it is both Markov and Gaussian. Here we abandon the Markov property but keep the Gaussianity. We assume a centered stationary Gaussian random process with finite energy. The statistical properties of such a process are fully determined by the covariance
\begin{equation}\label{correlatorgeneral}
\langle x(t) x(t^{\prime})\rangle = \mathcal{V} \,\kappa(t-t^{\prime}),
\end{equation}
where $\kappa(z)$, an even function of its argument, is normalized to unity,
\begin{equation}\label{normalization}
\int_{-\infty}^{\infty} \kappa(z) dz =1,
\end{equation}
and $\mathcal{V}>0$ is the process' magnitude. A convenient alternative is to define the process by its spectral density $\mathcal{V} \kappa_{\omega}$, where $\kappa_{\omega}$ is the Fourier transform of $\kappa(z)$:
\begin{equation}\label{kappak}
\kappa_{\omega} = \frac{1}{2 \pi} \int_{-\infty}^{\infty} e^{-i \omega z} \kappa(z) \,dz .
\end{equation}
$\kappa_{\omega}$ is a real function because of the symmetry $\kappa(-z)=\kappa(z)$, and we will assume that it is non-negative.

\begin{figure} [h]
\includegraphics[width=0.3\textwidth,clip=]{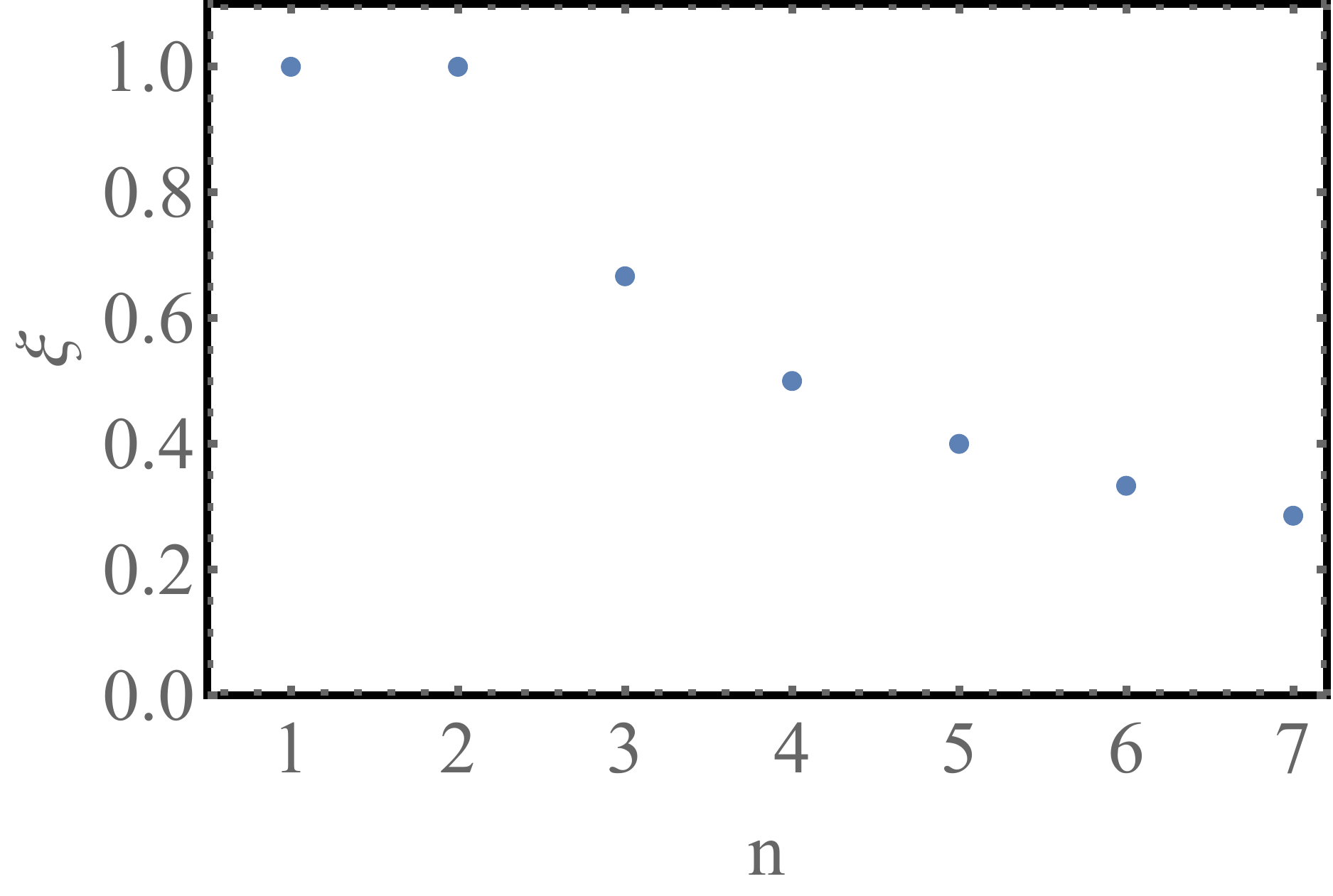}
\caption{The $n$-dependence of the exponent $\xi$ in the long-time large-$a$ expression $-\ln \mathcal{P}(a\to \infty,T\to \infty) \simeq T^{\xi} f(a)$.}
\label{intermittency}
\end{figure}

As in Ref. \cite{Nickelsen}, we will employ the OFM which correctly predicts the large-$a$ asymptotic of $-\ln \mathcal{P}(a,T)$ at any fixed $\mathcal{V}$. For non-Markov processes, that we are interested in, the OFM minimization procedure will lead us to a \emph{non-local} theory, in contrast to the local ``classical mechanics" of Ref. \cite{Nickelsen}.  Still, we will argue that the main predictions of Ref. \cite{Nickelsen} hold. That is, for $n=1$ and $2$  the normal scaling (\ref{normal}) is observed, as we obtain
\begin{equation}\label{actionfixedpoint}
-\ln \mathcal{P}(a \to \infty,T \to \infty) \simeq \frac{T a^{2/n}}{2\mathcal{V}}.
\end{equation}
At $n>2$  the normal scaling gives way to the anomalous scaling (\ref{normal}) with $\xi =2/n$. In this regime we obtain
\begin{equation}\label{generalscaling}
-\ln \mathcal{P}(a \to \infty,T \to \infty)\simeq \frac{C_n \tau^{1-\frac{2}{n}}A^{\frac{2}{n}}}{\mathcal{V}} ,
\end{equation}
where $A=aT$ is the area under the graph of $x^n(t)$. The dimensionless factor  $C_n$ depends on $n$ and on the problem-specific covariance $\kappa(z)$, but the scaling with $A$, $\mathcal{V}$ and $\tau$ is universal. Overall , the $n$-dependence of the exponent $\xi$ is (see Fig. \ref{intermittency})
$$
\xi =
\begin{cases}
1,   & \quad n=1, 2\,. \\
2/n,  & \quad n=3,4,5 \,, \dots .
\end{cases}
$$

The case of $n=2$ (when the scaling is normal) was solved by Bryc and Dembo exactly (that is, for any $a$) for arbitrary $\kappa_{\omega}$ \cite{BrycDembo}. We will compare our large-$a$ asymptotic~(\ref{actionfixedpoint}) with their exact result as we proceed.

\section{Optimal fluctuation method and calculations}
\label{main}

Our starting point is the statistical weight of a given realization of  the stationary Gaussian process $x(t)$, see \textit{e.g.} Ref. \cite{Zinn}. Up to normalization, the statistical weight is equal to $\sim \exp\{-S[x(t)]\}$,  where the action functional is \cite{finiteT}
\begin{equation}\label{actiongeneral}
S[x(t)] = \frac{1}{2\mathcal{V}} \int_{-\infty}^{\infty} dt \int_{-\infty}^{\infty} dt^{\prime} K(t-t^\prime) x(t) x(t^\prime).
\end{equation}
Here $K(t-t^{\prime})= K(t^{\prime}-t)$ is the inverse kernel, defined by the relation
\begin{equation}\label{inverse}
\int_{-\infty}^{\infty} dt^{\prime\prime} \,K(t-t^{\prime\prime}) \,\kappa(t^{\prime}-t^{\prime\prime}) =\delta(t-t^{\prime}).
\end{equation}
By virtue of Eq.~(\ref{inverse}), $K(\dots)$ is automatically normalized to unity: $\int_{-\infty}^{\infty} K(z) dz =1$. In addition, the Fourier transform of $K(z)$, which we will denote by $K_{\omega}$,  is simply related to $\kappa_{\omega}$: $K_{\omega}\kappa_{\omega} =(2\pi)^{-2}$.

Now we assume a very large $a$ and employ the OFM. We should minimize  the action functional~(\ref{actiongeneral})  over all possible paths $x(t)$ obeying the constraint $\bar{x^n}=a$.
Introducing a Lagrange multiplier, we proceed to minimize the modified functional
\begin{widetext}
\begin{equation}\label{functional}
S_{\lambda}[x(t)]= \frac{1}{2\mathcal{V}}\int_{-\infty}^{\infty} dt \left[\int_{-\infty}^{\infty} dt^\prime  K(t-t^\prime) x(t) x(t^\prime) -2\lambda x^n(t) \right] .
\end{equation}
A linear variation $\delta S_{\lambda}$ must vanish:
\begin{equation}\label{B90}
\delta S_{\lambda} = \frac{1}{\mathcal{V}}\int_{-\infty}^{\infty} dt \,\delta x(t) \left[\int_{-\infty}^{\infty} dt^{\prime} K(t-t^\prime) x(t^\prime)-\lambda n x^{n-1} (t)\right] =0.
\end{equation}
\end{widetext}
This condition generates a nonlocal theory, described by the integral equation
\begin{equation}\label{C100}
\int_{-\infty}^{\infty} dt^\prime\,K(t-t^\prime)x(t^\prime) = \lambda n x^{n-1}(t) .
\end{equation}
Using Eq. (\ref{inverse}), we can invert Eq.~(\ref{C100}) and arrive at an equivalent but more convenient equation with a well-behaved kernel $\kappa(z)$:
\begin{equation}\label{C105}
\lambda n \int_{-\infty}^{\infty} dt^\prime\,\kappa(t-t^\prime)x^{n-1}(t^\prime) = x(t) .
\end{equation}
Once Eq.~(\ref{C105}) is solved for $x(t)$ for a given $\lambda$, $\lambda$ should be expressed through $a$ from the condition $\bar{x^n}=a$.

For any $n$ Eqs.~(\ref{C100}) and~(\ref{C105}) have a constant solution, $x(t)=\text{const}$. For $\lambda \equiv \lambda_n = (1/n) \,a^{\frac{2-n}{n}}$, the constant solutions, obeying the condition $\bar{x^n} =a$, are $x(t)=a^{1/n}$ for even $n$, and $x(t)=\pm a^{1/n}$ for odd $n$.  Let us consider the cases of $n=1$, $n=2$ and $n>2$ in more detail.

\subsection{$n=1$ and $2$}

For $n=1$ Eq.~(\ref{C105}) degenerates into $x(t)=\lambda$, and we must choose  $\lambda=a$. Here $x=a$ is the only solution:
the system stays at the fixed point of our nonlocal ``classical theory". Now we can evaluate the action (\ref{actiongeneral}), or rather directly compute $f(a)$:
\begin{equation}\label{fageneral}
f(a) = \lim_{T\to \infty} \frac{a^2}{2\mathcal{V} T} \int_{-T/2}^{T/2} dt \int_{-T/2}^{T/2} dt^{\prime} K(t-t^\prime).
\end{equation}
The $T\to \infty$ limit of the internal integral is equal to unity, and we arrive at a simple large-$a$ asymptotic
\begin{equation}\label{fasimple}
f(a) = \frac{a^2}{2\mathcal{V}}\quad \text{for}\quad n=1,
\end{equation}
corresponding to a Gaussian tail of the distribution $\mathcal{P}(a,T)$,  for the whole class of stationary Gaussian processes that we consider. For the OU process the rate function (\ref{fasimple}) is exact, 
see \textit{e.g.} Ref. \cite{Touchette2018}.

For $n=2$ Eq.~(\ref{C105}) is a homogeneous linear Fredholm equation of the second kind, and $2\lambda$ plays the role of an eigenvalue. (An infinite number of) nonzero solutions exist only for $\lambda=1/2$, and all of them are constant. The correct constrained solutions, $x=\pm \sqrt{a}$, are set by the condition $\bar{x^2}=a$. This leads to
\begin{equation}\label{fbsimple}
f(a\to \infty)=\frac{a}{2 \mathcal{V}} \quad \text{for}\quad n=2,
\end{equation}
describing an exponential tail of $\mathcal{P}(a,T)$. Equation~(\ref{fbsimple}) agrees with the large-$a$ asymptotic of the \emph{exact} rate function $f(a)$ for $n=2$, obtained by Bryc and Dembo \cite{BrycDembo}.  Indeed, their $f(a)$ is given, in our notation, by the Fenchel-Legendre transform
$$
f(a) = \max_y \,\left[ay-L(y)\right]
$$
of the function
\begin{equation}\label{Ly}
L(y)=-\frac{1}{4\pi} \int_{-\infty}^{\infty} \ln \left(1-4\pi y \mathcal{V}\kappa_{\omega}\right) d\omega,
\end{equation}
where $-\infty<y<(2\mathcal{V})^{-1}$. For $a\to \infty$ the maximum of the function $\phi(y)=ay-L(y)$ is achieved at the maximum allowable value of $y$: $y=y_{\text{max}}=(2\mathcal{V})^{-1}$. Therefore, $\max \phi(y)\simeq a y_{\text{max}}$, whereas the function $L(y)$ does not contribute in the leading order, and we arrive at Eq.~(\ref{fbsimple}).

Equations~(\ref{fasimple}) and (\ref{fbsimple}) show that different Gaussian processes  with different correlations but  the same magnitude $\mathcal{V}$ have exactly the same large-$a$ asymptotics of the rate functions $f(a)$. The situation is different for processes with the same \emph{variance} $\text{Var} = \mathcal{V}\kappa(0)$, but different covariances.  As an example, let us consider a process with a non-monotonic covariance:
\begin{equation}\label{osccorrelator}
\langle x(t) x(t^{\prime})\rangle = \frac{\epsilon^2 \tau}{2}\,\,e^{-\frac{|t-t^\prime|}{\tau}}\, \cos\left[\Omega(t-t')\right],
\end{equation}
where there are alternating regions of positive and negative correlations. This process generalizes the OU process and reduces to the latter when $\Omega=0$. The process (\ref{osccorrelator}) has the same variance $\epsilon^2 \tau/2$ as the OU process (\ref{OULangevin}), but the
magnitude of the process (\ref{osccorrelator}), $\mathcal{V}=\epsilon^2 \tau^2 (1+\Omega^2 \tau^2)^{-1}$, is smaller than that of the OU process. As a result, the rate functions $f(a)$ of the process (\ref{osccorrelator}) for $n=1$ and $2$  are larger
(for $\Omega \tau \gg 1$ much larger) than those of the OU process. That is, the presence of negative correlations makes the observation of a given value of the long-time averages $\bar{x}$ and $\bar{x^2}$ less likely.

\subsection{$n>2$}

For $n>2$ the integral equation~(\ref{C100}) is nonlinear, and one can expect multiple solutions. When more than one solution is present, and they have different actions, the one with the least action must be selected. Before we continue, let us use  Eq.~(\ref{C105}) to simplify the expression (\ref{actiongeneral}) for the action. We can rewrite Eq.~(\ref{C105}) as
$$
x(t')=\lambda n \int_{-\infty}^{\infty} dt^{\prime\prime}\,\kappa(t'-t^{\prime\prime})x^{n-1}(t^{\prime\prime})
$$
and plug this expression into Eq.~(\ref{actiongeneral}). By virtue of Eq.~(\ref{inverse}), the integral over $t^{\prime}$  yields the delta-function $\delta(t-t^{\prime\prime})$. After integration over $t^{\prime\prime}$, we obtain a simple expression
\begin{equation}\label{Smixed}
S = \frac{\lambda n}{2\mathcal{V}} \int_{-\infty}^{\infty} x^n(t) dt = \frac{\lambda n T a}{2\mathcal{V}},
\end{equation}
which is valid for any $n$. We still need to express $\lambda$ through $a$ and $T$.
For the constant solution this leads to Eq.~(\ref{actionfixedpoint}),
reproducing Eqs.~(\ref{fasimple}) and ~(\ref{fbsimple}) for $n=1$ and $2$, respectively.

Now let us consider $n>2$ and assume that a localized solution exists. In order to determine the scaling behavior of $S \simeq -\ln \mathcal{P}(a,T)$, let us return to Eq.~(\ref{C105}) and introduce the dimensionless variable $y(t) = \lambda^{\frac{1}{n-2}} x(t)$.  The resulting equation for $y(t)$,
\begin{equation}\label{C110}
n \int_{-\infty}^{\infty} dt^\prime\,\kappa(t-t^\prime)y^{n-1}(t^\prime) = y(t) ,
\end{equation}
is $\lambda$-independent. Its solution $y(t)$ should obey the constraint $\bar{x^n}=a$, and we obtain
\begin{equation}\label{bT}
\int_{-\infty}^{\infty} x^n(t) dt = \lambda^{\frac{n}{2-n}} \int_{-\infty}^{\infty} y^n(t) dt =a T,
\end{equation}
assuming that the integral converges. The quantity $aT\equiv A$ is the area under the graph of $x^n(t)$. As we can see, the $A$-distribution is $T$-independent at large $T$, as a consequence of the localization of the optimal path $x(t)$ on the time scale of the correlation time $\tau$. Rescaling time by the correlation time $\tau$, $\tilde{t}=t/\tau$, we obtain from
Eq.~(\ref{bT})
\begin{equation}\label{lambdageneral}
\lambda =B_n  \left(\frac{\tau}{A}\right)^{\frac{n-2}{n}},
\end{equation}
where the dimensionless factor
\begin{equation}\label{Cn}
B_n=\left[\int_{-\infty}^{\infty} y^n(\tilde{t}) \, d\tilde{t}\,\right]^{\frac{n-2}{n}}
\end{equation}
depends only on $n$ and on the particular form of the covariance.
Plugging Eq.~(\ref{lambdageneral}) into Eq.~(\ref{Smixed}), we arrive at the announced Eq.~(\ref{generalscaling}) with $C_n=(n/2)B_n$. Comparing Eqs.~(\ref{generalscaling}) and~(\ref{actionfixedpoint}), we see that a localized solution, when it exists, provides a lesser action than the constant solution and should therefore be selected.

Integrating both parts of Eq.~(\ref{C110}) over $t$ and using Eq.~(\ref{normalization}), we see that the localized solutions obey the general relation
\begin{equation}\label{momentsrelation}
\int_{-\infty}^{\infty} y(t) dt = n \int_{-\infty}^{\infty} y^{n-1}(t) dt, \quad n>2.
\end{equation}

Now we consider some examples of Gaussian processes, where all the calculations can be performed analytically, demonstrating the existence of localized solutions and anomalous scaling of $\ln \mathcal{P}(a,T)$ at $n>2$.

\subsection{Analytical solutions}

\subsubsection{The OU process}
\label{OUexample}

We start by revisiting the anomalous scaling of the OU process,  see Eqs. (\ref{OULangevin}) and (\ref{OUcorrelator}), studied by NT \cite{Nickelsen}.
Here $\kappa(z)= (2\tau)^{-1} e^{-|z|/\tau}$ and $\mathcal{V}=\epsilon^2 \tau^2$.  The spectral density is
\begin{equation}\label{kappakOU}
\kappa_{\omega}=
\frac{1}{2\pi (1+\omega^2 \tau^2)}.
\end{equation}
Using the relation  $K_{\omega}\kappa_{\omega} =(2\pi)^{-2}$, we obtain
\begin{equation}\label{KkOU}
K_{\omega} = \frac{1+\omega^2\tau^2}{2\pi}.
\end{equation}
Expressing $K(z)$ in Eq.~(\ref{actiongeneral}) as the inverse Fourier transform of this $K_{\omega}$,  we obtain after some algebra
\begin{equation}\label{action1}
S[x(t)] = \frac{1}{2\epsilon^2} \int_{-\infty}^{\infty} dt\,\left(\dot{x}^2+\frac{x^2}{\tau^2}\right),
\end{equation}
the familiar action functional of the OU process. Furthermore, using the inverse Fourier transform, we can recast the integral equation~(\ref{C100}) into a second-order ordinary differential equation:
\begin{equation}\label{odeOU}
\ddot{x}(t)-\frac{x(t)}{\tau^2}+\frac{\lambda n}{\tau^2}\,x^{n-1}(t) =0 ,
\end{equation}
which is nothing but the Euler-Lagrange equation for the action~(\ref{action1}), with the constraint $\bar{x^n} = a$ accommodated via a Lagrange multiplier \cite{Nickelsen}. NT determined the action, corresponding to the (zero-energy) homoclinic solution of Eq.~(\ref{odeOU}) for $n>2$, without finding the solution itself. For our purposes we need the homoclinic solution, and it can be found in a straightforward manner:
\begin{equation}\label{OUxt}
 x(t) = (2\lambda)^{\frac{1}{2-n}} \,\text{sech}^{\frac{2}{n-2}} \left[\frac{(n-2)t}{2\tau}\right] ,
\end{equation}
up to an arbitrary time shift, $t \to t+C$ \cite{shift}. As one can see, the optimal path is exponentially localized in time in this example. A direct integration shows that this $x(t)$  solves our integral equation~(\ref{C105})  with $\kappa(z)= (2\tau)^{-1} e^{-|z|/\tau}$ . The Lagrange multiplier $\lambda$ is given by Eqs.~(\ref{lambdageneral}) and (\ref{Cn}). The action, calculated from Eq.~(\ref{Smixed}),  conforms to the general scaling form~(\ref{generalscaling}) and coincides with the action found by NT \cite{Nickelsen}.

\subsubsection{Gaussian covariance}
\label{gauss}

As a previously unexplored example, let us consider a Gaussian covariance:
\begin{equation}\label{correlatorgauss}
\langle x(t) x(t^\prime)\rangle = v \,e^{-(t-t^\prime)^2/\tau^2},
\end{equation}
with variance $v$ and the correlation time $\tau$. In this case $\kappa(z)=(\sqrt{\pi} \tau)^{-1} e^{-z^2/\tau^2}$ and $\mathcal{V}=\sqrt{\pi} \tau v$. This process is non-Markov, so there is no local stochastic ODE that would describe it. But here too there is a localized solution $x(t)$ of  Eq.~(\ref{C105}) as soon as $n>2$. This solution can be easily guessed to be a Gaussian:
\begin{equation}\label{simplegauss}
x(t) = \beta_n e^{-\frac{t^2}{\sigma_n^2}},
\end{equation}
up to an arbitrary time shift.  We plug the Ansatz~(\ref{simplegauss}) into Eq.~(\ref{C105}) and determine the \textit{a priori} unknown $\beta_n$ and $\sigma_n$. The result, in terms of $\lambda$, is
\begin{equation}\label{xtgauss}
x(t)=\left(\frac{\sqrt{n-1}}{\lambda
   n}\right)^{\frac{1}{n-2}}
   e^{-\frac{(n-2) t^2}{(n-1) \tau ^2}},\quad n>2.
\end{equation}
Here the solution is localized even stronger than exponentially. Again, Eqs.~(\ref{lambdageneral}) and (\ref{Cn}) give $\lambda$ and,
using Eq.~(\ref{Smixed}), we finally obtain
\begin{equation}\label{actiongaussian}
S= \frac{c_n}{v} \left(\frac{A}{\tau}\right)^{2/n},
\end{equation}
where
$$
c_n = \frac{(n-1)^{\frac{n-1}{n}}}{2 \pi^{\frac{1}{n}}\left[n (n-2)\right]^{\frac{n-2}{2n}}}.
$$
This result conforms to the general scaling form~(\ref{generalscaling}).

\subsubsection{Inverse problem}
We are unaware of a general method of solving the nonlinear integral equation~(\ref{C110}) analytically for a given $\kappa(z)$. Many instructive examples, however,  can be produced by solving the \emph{inverse} problem,  that is by determining the covariance $\kappa(z)$, for which the equation has
a specified localized solution. Let this solution be $y(t)=\alpha f(t)$, where $f(0)=1$, and $\alpha>0$ is an a priori unknown constant. We demand that the Fourier transforms of $f(t)$ and $f^{n-1}(t)$, that we will denote
by $f_{\omega}$ and $(f^{n-1})_{\omega}$, exist and are positive.  An additional condition will arise shortly.     Using the convolution theorem, we can rewrite Eq.~(\ref{C110}) as
\begin{equation}\label{kappaomega1}
\kappa_{\omega} = \frac{f_{\omega}}{2\pi n \alpha^{n-2} \left(f^{n-1}\right)_{\omega}}.
\end{equation}
In view of the normalization condition (\ref{normalization}) we must demand $\kappa_{\omega=0}=(2\pi)^{-1}$, which sets $\alpha$:
\begin{equation}\label{alpha}
\alpha^{n-2} = \frac{f_{\omega=0}}{n \left(f^{n-1}\right)_{\omega=0}}.
\end{equation}
As a result,
\begin{equation}\label{kappaomega}
\kappa_{\omega} = \frac{1}{2\pi}\,\frac{f_{\omega} \left(f^{n-1}\right)_{\omega=0}}{f_{\omega=0}\left(f^{n-1}\right)_{\omega}}.
\end{equation}
If this $\kappa_{\omega}$ obeys the additional condition $\int_{-\infty}^{\infty} \kappa_{\omega} d\omega < \infty$,
which can be written as
\begin{equation}\label{boundedintegral}
\int_{-\infty}^{\infty} \frac{f_{\omega}}{\left(f^{n-1}\right)_{\omega}} d\omega < \infty,
\end{equation}
the desired covariance $\kappa(z)$ of our stationary Gaussian process exists and is given by the inverse Fourier transform of $\kappa_{\omega}$. Now we return from $y(t)$ to $x(t)$, express $\lambda$ through $A$ from the condition $\bar{x^n}=a$, and calculate the action. Using Eqs.~(\ref{Smixed}) and~(\ref{alpha}), we finally arrive at Eq.~(\ref{generalscaling}), where
\begin{equation}\label{Cninverse}
C_n= \frac{f_{\omega=0}}
{2 \left[\int\limits_{-\infty}^{\infty} f^n(z) dz\right]^{\frac{2}{n}-1}\left(f^{n-1}\right)_{\omega=0}}.
\end{equation}

\begin{figure}[ht]
\includegraphics[width=0.3\textwidth,clip=]{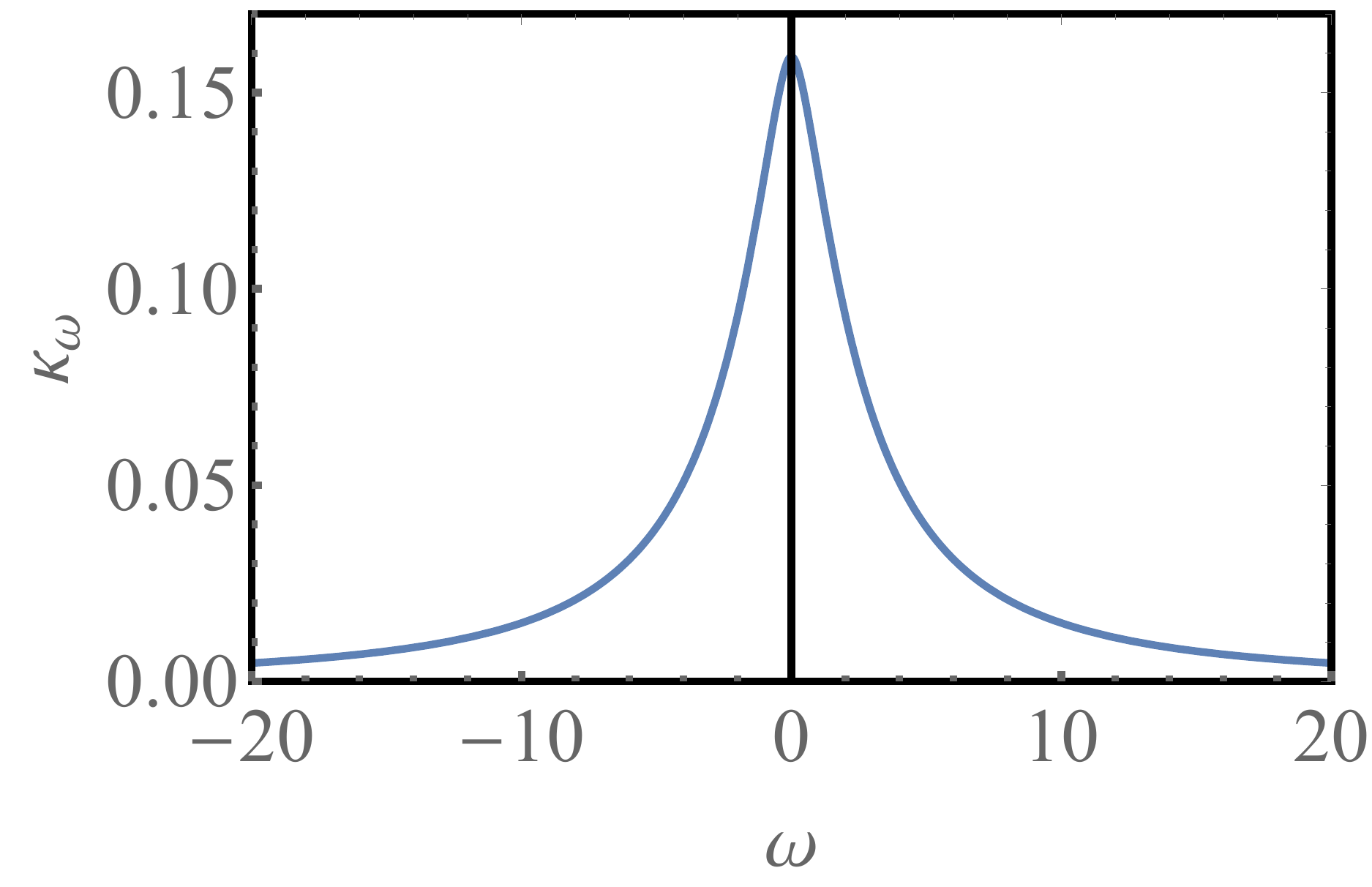}
\includegraphics[width=0.3\textwidth,clip=]{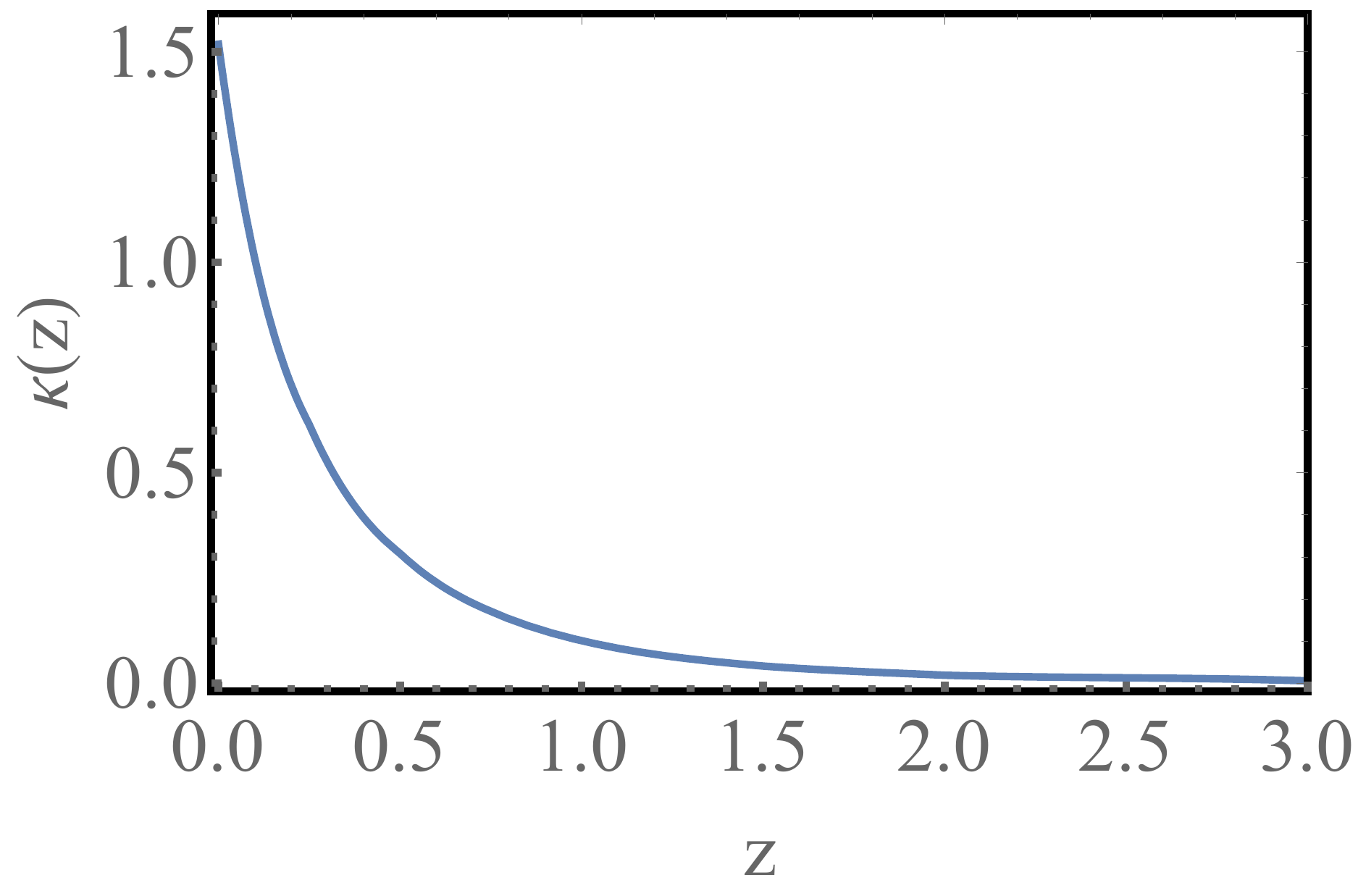}
\caption{The spectral density (\ref{kappaomegaexample}) (top) and the covariance $\kappa(z)$ given by its inverse Fourier transform
(bottom). In this case the (rescaled) localized solution for $n=3$ is $y(t)=(8/15) (1+t^2)^{-2}$.}
\label{engineered}
\end{figure}

Here is one of many exactly solvable examples that can be obtained using this method. Let $n=3$. Which covariance $\kappa(z)$ gives rise to the localized solution $y(t)=\alpha (1+t^2/\tau^2)^{-2}$
as the optimal path conditioned on $\bar{x^3}=a$? In this example the localization of the optimal solution is only algebraic. Let us set $\tau=1$  for brevity. Then $f(t)=(1+t^2)^{-2}$, and the Fourier transforms $f_{\omega}$ and $(f^2)_{\omega}$,
\begin{eqnarray}
\label{twofourier}
  f_{\omega} &=& \frac{1}{4} e^{-|\omega|} \left(1+|\omega|\right) ,  \\
  (f^2)_{\omega} &=& \frac{1}{96} e^{-|\omega|} \left(15+15|\omega|+6\omega^2+|\omega|^3\right),
\end{eqnarray}
are everywhere positive. Equation~(\ref{kappaomega}) yields the spectral density
\begin{equation}\label{kappaomegaexample}
\kappa_{\omega}=\frac{15\, (\left| \omega\right| +1)}{2 \pi  \left(\left|
   \omega\right|^3+6 \omega^2+15 \left| \omega\right|
   +15\right)},
\end{equation}
which decays as $\omega^{-2}$ as $|\omega| \to \infty$ and therefore satisfies the finite-energy condition~(\ref{boundedintegral}). This spectral density and the resulting covariance $\kappa(z)$, for $z>0$, are shown in Fig. \ref{engineered}.

\begin{figure}
\includegraphics[width=0.3\textwidth,clip=]{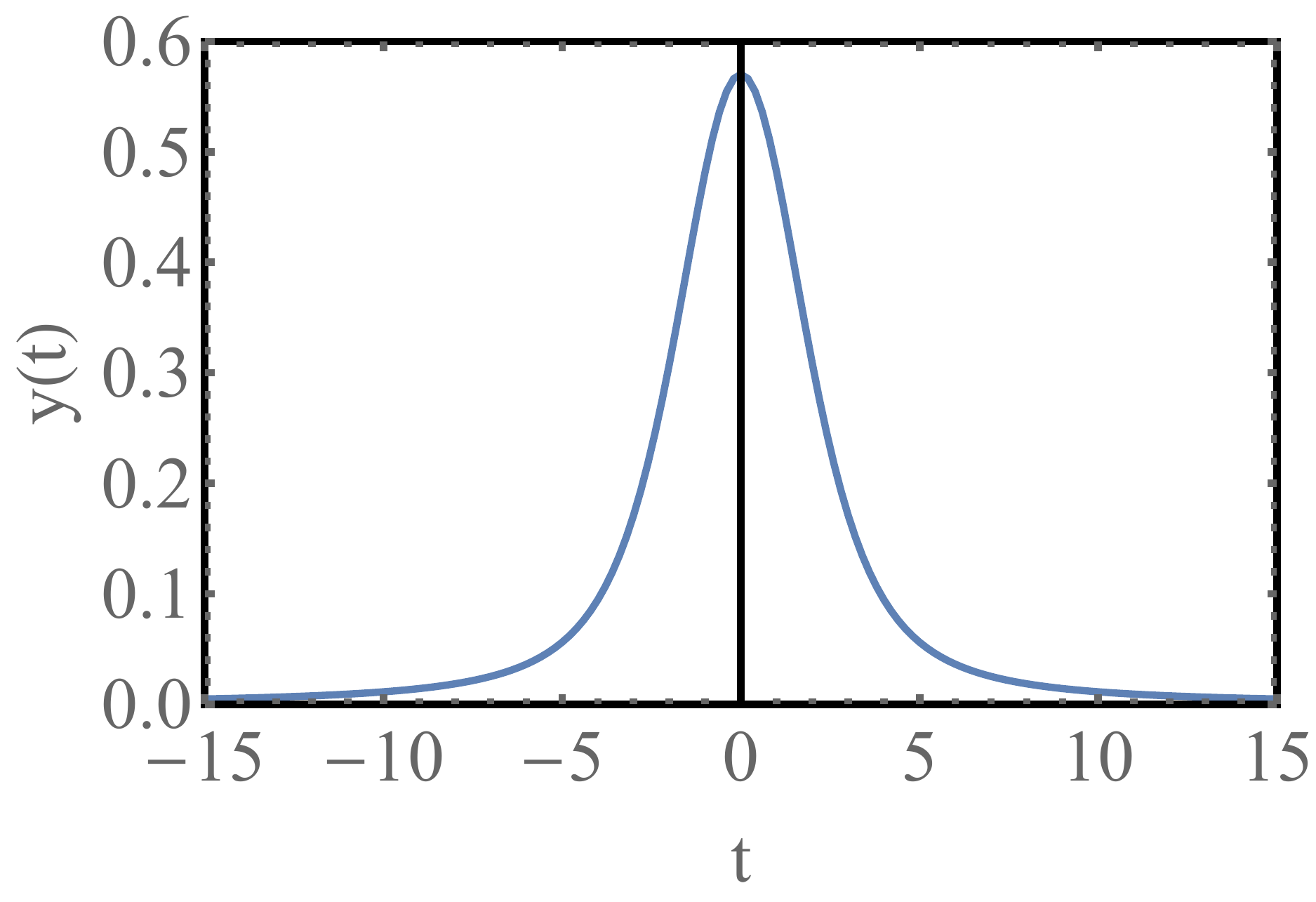}
\caption{The localized solution of Eq.~(\ref{C110}) for $n=3$, obtained numerically for $\kappa(z)=(\pi \tau)^{-1} (1+z^2/\tau^2)^{-1}$. This solution decays as $t^{-2}$ at large $|t|$. We set $\tau=1$.}
\label{ynumeric}
\end{figure}

\subsection{Numerical solution}

For a general $\kappa(z)$ the localized solutions can be found by solving the integral equation~(\ref{C110}) numerically for specified $n$.
We found it convenient to discretize the integral in the left hand side of Eq.~(\ref{C110}) in a straightforward way and use the standard \emph{FindRoot} option of ``Mathematica" to solve the resulting system of nonlinear algebraic equations.
The success of numerical solution depends on the choice of the (localized) trial function. For a wrong choice ``Mathematica" returns a trivial, constant (up to boundary layers of numerical origin) or more complicated, oscillating solution. For suitable functions $\kappa(z)$ these may be correct solutions of Eq.~(\ref{C110}), but they are non-optimal \cite{compareOU}. Changing the amplitude and width of the localized trial function, we obtained well-behaved localized solutions for many different $\kappa(z)$. We tested the accuracy of the numerical method on the two exactly solvable solutions from Secs. \ref{OUexample} and \ref{gauss}, and observed a very good accuracy. With the numerical solution at hand, one can evaluate the constant $C_n$ in  the general expression~(\ref{generalscaling}).  Figure \ref{ynumeric}
shows the numerical localized solution $y(t)$ in one of the examples that we explored: for $n=3$  and the covariance
$\kappa(z)=(\pi \tau)^{-1} (1+z^2/\tau^2)^{-1}$.

\section{Discussion}

We demonstrated that an anomalous scaling with time of $\ln \mathcal{P}(a,T\to \infty)$ at $n>2$ holds for a whole class of stationary Gaussian processes. The anomalous scaling, that we probed at $a\to \infty$, is closely related to the existence of a localized solution
of the nonlinear integral equation (\ref{C110}). This solution describes, at $n>2$, the most probable trajectory $x(t)$, conditioned on the area under $x^n(t)$.  A natural conjecture is that a localized solution exists if the spectral density $\kappa_{\omega}$ of the Gaussian process in question is  bounded,  positive and has a finite energy: $\int_{-\infty}^{\infty} \kappa_{\omega} d\omega < \infty$. It would be very interesting to prove (or improve) this conjecture.  Finally, it is both challenging and important to devise a method that would allow one to go beyong the large-$a$ asymptotics (\ref{actionfixedpoint}) and (\ref{generalscaling}) and calculate  $\ln \mathcal{P}(a,T\to \infty)$ exactly in the anomalous scaling regime $n>2$. The most general scaling behavior of $\ln \mathcal{P}$ at long times can be represented as $-\ln \mathcal{P}(a,T\to \infty) = T^{\mu} \phi(A/T^{\nu},n)$. The large-$a$ asymptotic (\ref{generalscaling}) imposes a relation between the presently unknown exponents $\mu$ and $\nu$: $\nu = n \mu/2$, and we obtain
\begin{equation}\label{singleexponent}
-\ln \mathcal{P}(a,T\to \infty) = T^{\mu} \phi \left(\frac{A}{T^{\frac{n\mu}{2}}},n\right),
\end{equation}
leaving us with a single exponent $\mu$ to be found.

\bigskip
\section*{Acknowledgments}

I am grateful to Tal Agranov, Pavel Sasorov and Hugo Touchette for useful discussions, and to Hugo Touchette for a critical reading of the manuscript. This work was supported by the Israel
Science Foundation (Grant No. 807/16).


\begin{thebibliography}{99}
\bibitem{finiteT} For the localized solutions, corresponding to the anomalous scaling, the integration limits can be set $-\infty$ and $\infty$. For the constant solutions, corresponding to the normal scaling, we are actually interested in
    $$
    f(a) = \lim_{T\to \infty} \frac{1}{2\mathcal{V} T} \int_{-T/2}^{T/2} dt \int_{-T/2}^{T/2} dt^{\prime}  K(t-t^\prime) x(t) x(t^{\prime}).
    $$
\end{thebibliography}
\end{document}